\documentclass[twoside,fleqn]{article}
\usepackage{espcrc2}
\usepackage{amssymb,amsmath}

\usepackage{graphicx}
\DeclareGraphicsExtensions{.jpg,.jpeg,.eps.ps}

\newcommand{\pslash}{\not\!p}
\newcommand{\qslash}{\not\!q}

\newcommand{\kslash}{\not\!k}
\newcommand{\Pslash}{\not\!P}
\newcommand{\half}{\frac{1}{2}}

\newcommand{\Tr}{\operatorname{tr}}
\newcommand{\order}{\mathcal{O}}

\newcommand{\gm}{\gamma_\mu}

\newcommand{\smn}{\sigma_{\mu\nu}}
\newcommand{\snl}{\sigma_{\nu\lambda}}

\newlength{\colw}
\newcommand{\cgluons}{\cite{Bonnet:2001uh,Bowman:2004jm}}
\newcommand{\cqqg}{\cite{Skullerud:2002ge}}
\newcommand{\cstruct}{\cite{Skullerud:2003qu}}

\title{Quark--gluon vertex in arbitrary kinematics\thanks{Talk by
    JIS at QCD Down Under, Adelaide, Australia, 10--19 March 2004.}}

\author{Jon-Ivar Skullerud\address[TCD]{School of Mathematics, 
    Trinity College, Dublin 2, Ireland},
  Patrick O.\ Bowman\address[Ind]{Nuclear Theory Center,
Indiana University, Bloomington IN 47408, USA}
  Ay{\c s}e K{\i}z{\i}lers\"u\address[CSSM]{Centre for the Subatomic
    Structure of Matter, Adelaide University, Adelaide, SA 5005,
    Australia}, Derek B.\ Leinweber\addressmark[CSSM],
  Anthony G.\ Williams\addressmark[CSSM]}

\begin{document}
%%%% hack to get flush-left equations with zero indent in amsmath
\makeatletter \@mathmargin = 0pt \makeatother
%%%%
%\bibliographystyle{h-physrev4}

\begin{abstract}
We compute the quark--gluon vertex in quenched lattice QCD, in the
Landau gauge using an off-shell mean-field $\order(a)$-improved
fermion action.  The complete vertex is computed in two specific
kinematical limits, while the Dirac-vector part is computed for
arbitrary kinematics. We find a nontrivial and rich tensor structure,
including a substantial infrared enhancement of the interaction
strength regardless of kinematics.
\end{abstract}

\maketitle

\section{INTRODUCTION}

Over the past few years, substantial progress has been made in our
understanding of the nonperturbative correlation functions
(propagators and vertices) of the fundamental fields of QCD
and their relation to the phenomena of colour
confinement and dynamical chiral symmetry breaking.  It has recently
become evident that at least in Landau gauge, a detailed knowledge of
the structure of the quark--gluon vertex is essential for an
understanding of the dynamics of quark confinement and chiral symmetry
breaking, which is encoded in the quark Dyson--Schwinger equation
(DSE), relating
the quark propagator $S(p)$ to the gluon propagator and the
quark--gluon vertex $\Gamma_\mu(p,q)$, where $p$ and $q$ are quark and
gluon momenta respectively.
%
%\begin{multline}
%%\begin{equation} 
%%\begin{split}
%S^{-1}(p) = i\pslash + m \\
% + \frac{4g^2}{3}\!\int\!\!\frac{d^4q}{(2\pi)^4}  
%\gamma_\mu S(p\!-\!q)D_{\mu\nu}(q)\Gamma_\nu(p\!-\!q,q) \, .
%%\end{split}
%\label{eq:quark-dse} 
%%\end{equation} 
%\end{multline}
%
%Here, $\Lambda_\nu(p,q)=ig\Gamma_\nu(p,q)$ is the quark--gluon vertex,
%while $D_{\mu\nu}(q)=P_{\mu\nu}(q)D(q^2)$ is the gluon propagator,
%with $P_{\mu\nu}(q)$ the transverse projection operator.

The overall shape of the gluon propagator is now quite well
known, both from lattice QCD \cgluons\ and from studies of the coupled
ghost--gluon Dyson--Schwinger equations.  It is now clear that if this
is fed into the quark DSE together with a bare or QED-like
vertex, the resulting quark propagator will not
exhibit a sufficient degree of chiral symmetry breaking.  Several of
the contributions at this conference have addressed this issue, in
various ways \cite{Holl:2004qn,Iida:2003xe,Fischer:2004ym,Tandy:2004rk}.

The quark--gluon vertex is related to the ghost sector through the
Slavnov--Taylor identity (STI),
\begin{equation}
\begin{split}
q^\mu\Gamma_\mu(p,q) &= G(q^2)
 \Bigl[(1-B(q,p+q))S^{-1}(p)\\
& - S^{-1}(p+q)(1-B(q,p+q))\Bigr] \, ,
\end{split}
\label{eq:sti}
\end{equation}
where $G(q^2)$ is the ghost renormalisation function and $B^a(q,k)$ is
the ghost--quark scattering kernel.  In particular, if the ghost
propagator is infrared enhanced, as both lattice \cite{Bloch:2003sk}
and DSE studies \cite{Fischer:2002hn} indicate, the vertex will also
be so.  This provides for a consistent picture of confinement and
chiral symmetry breaking at the level of the Green's functions of
Landau-gauge QCD, where the same infrared enhancement that is
responsible for confinement of gluons, provides the necessary
interaction strength to give rise to dynamical chiral symmetry
breaking in the quark sector.

Confinement of quarks is still not understood in this picture,
however.  If the effective interaction between a quark and an
antiquark by way of exchange of a nonperturbative gluon is to give
rise to a linearly confining potential, the quark--gluon vertex must
contain an infrared enhancement over and above that contained in the
ghost self-energy.  In the STI, this would be encoded in a non-trivial
ghost--quark scattering kernel.  Confirming or refuting this picture
is a major challenge for current lattice and DSE studies of
Landau-gauge QCD.

Another area where the quark--gluon vertex may be of interest is that
of effective charges.  Although `the running coupling' is not a
meaningful concept beyond perturbation theory, since there is no known
way of nonperturbatively connecting two different `schemes',
process-dependent effective charges may be defined non-perturbatively
and be phenomenologically useful.  The interaction between quarks and
gluons may be a natural starting point for many of the physically
interesting processes.

Here we will present results of a lattice investigation into the
quark--gluon vertex \cqqg.  This consists of a determination of the
full structure of the vertex at two particular kinematical points (the
{\em soft gluon point} where the gluon has zero momentum, and the {\em
quark reflection point} where the incoming and outgoing quark momenta
are equal and opposite), as well as a determination of the dominant,
vector part of the vertex in arbitrary kinematics.

\section{FORMALISM}

We denote the outgoing quark momentum $p$ and the outgoing gluon
momentum $q$.  The incoming quark momentum is $k=p+q$.
In the continuum, the quark--gluon vertex can be decomposed into four
components $L_i$ contributing to the Slavnov--Taylor identity and
eight purely transverse components $T_i$:
\begin{equation} 
\begin{split}
\Gamma_\mu(p,q) & = 
 \sum_{i=1}^{4}\lambda_i(p^2,q^2,k^2)L_{i,\mu}(p,q) \\
 &\phantom{=} + \sum_{i=1}^{8}\tau_i(p^2,q^2,k^2)T_{i,\mu}(p,q) \, .
\end{split}
\label{eq:decomp}
\end{equation}
In euclidean space, the components $L_i$ and $T_i$ are given by \cqqg
\small
\begin{alignat}{2}
L_{1,\mu}  =& \gamma_\mu &\qquad
L_{2,\mu}  =& -\Pslash P_\mu \\
L_{3,\mu}  =& -iP_\mu &\qquad 
L_{4,\mu}  =& -i\sigma_{\mu\nu}P_\nu \notag \\
T_{1,\mu}  =& -i\ell_\mu &\qquad
T_{2,\mu}  =& -\Pslash\ell_\mu \notag\\
T_{3,\mu}  =& \qslash q_\mu - q^2\gamma_\mu \notag \\
T_{4,\mu}  =& -i\bigl[q^2\smn P_\nu& + 2q_\mu\snl p_\nu &k_\lambda\bigr]
  \notag\\
T_{5,\mu}  =& -i\sigma_{\mu\nu}q_\nu &\qquad
T_{6,\mu}  =& (qP)\gamma_\mu - \qslash P_\mu\!\! \\
T_{7,\mu}  =& -\frac{i}{2} (qP)\smn& P_\nu
 - iP_\mu\snl& p_\nu k_\lambda \notag\\
T_{8,\mu}  =& -\gamma_\mu\sigma_{\nu\lambda}p_\nu k_\lambda&
 - \pslash k_\mu + \kslash p_\mu& \notag
\end{alignat}
\normalsize
where $P_\mu\equiv p_\mu+k_\mu$, $\ell_\mu\equiv
(pq)k_\mu-(kq) p_\mu$.  In Landau gauge, for $q\neq0$, we are only
able to compute the transverse projection of the vertex,
$\Gamma^P_\mu(p,q) \equiv P_{\mu\nu}(q)\Gamma_\nu(p,q)$, where
$P_{\mu\nu}(q) \equiv \delta_{\mu\nu}-q_{\mu}q_{\nu}/q^2$ is the
transverse projector.  Since the vertex will always be coupled to a
gluon propagator which contains the same projector, this is also the
only combination that appears in any application.  The four functions
$L_{i,\mu}$ are projected onto the transverse $T_{i,\mu}$, giving
rise to modified form factors
\begin{alignat}{2}
\lambda'_1 &= \lambda_1 - q^2\tau_3 &\, ; \qquad
\lambda'_2 &= \lambda_2 - \frac{q^2}{2}\tau_2 \, ;\\
\lambda'_3 &= \lambda_3 - \frac{q^2}{2}\tau_1 & \, ;\qquad
\lambda'_4 &= \lambda_4 + q^2\tau_4 \, .\notag
\end{alignat}
The lattice tensor structure is more complex, and
(\ref{eq:decomp}) is only recovered in the continuum.  The form
factors also receive large contributions from lattice artefacts at
tree level; the procedure we apply in correcting for these is
described in \cstruct.

In QED, the four form
factors $\lambda_i$ are completely determined by the fermion
propagator $S^{-1}(p) = i\pslash A(p^2) + B(p^2)$:
\begin{align}
\lambda_1(p^2,q^2,k^2) &= \half\left(A(p^2)+A(k^2)\right)\, ;
\label{eq:bc1} \\
\lambda_2(p^2,q^2,k^2) &= -\half\frac{A(p^2)-A(k^2)}{p^2-k^2} \, ;
\label{eq:bc2} \\
\lambda_3(p^2,q^2,k^2) &= \frac{B(p^2)-B(k^2)}{p^2-k^2} \, ;
\label{eq:bc3} \\
\lambda_4(p^2,q^2,k^2) &= 0 \, .
\label{eq:bc4}
\end{align}
By comparing our results with these forms we can get an idea of the
importance of the nonabelian contributions to the STI, and in
particular the quark--ghost scattering kernel.

\section{RESULTS}

We have analysed 495 configurations on a $16^3\times48$ lattice at
$\beta=6.0$, using a mean-field improved SW action with a quark mass
$m\approx115$ MeV.  This is part of the UKQCD data set described in
\cite{Bowler:1999ae}; further details can also be found in
\cite{Skullerud:2002ge}.  A second quark mass $m\approx60$ MeV has
also been studied, but as the mass dependence was found to be almost
negligible \cstruct, we do not show those results here.

\subsection{Soft gluon point $q=0$}

At $q=0$ the vertex reduces to
\begin{equation}
\begin{split}
\Gamma_\mu(p,0) =& \lambda_1(p^2)\gm \\
& - 4\lambda_2(p^2)\pslash p_\mu 
 - 2i\lambda_3(p^2)p_\mu \, ,
\end{split}
\label{eq:decomp-asym}
\end{equation}
where for brevity we write $\lambda_i(p^2,0,p^2)=\lambda_i(p^2)$.  In
this specific kinematics, the QED expressions
(\ref{eq:bc1})--(\ref{eq:bc3}) become\footnote{In \cstruct\ the
expressions for $\lambda_2$ and $\lambda_3$ had the wrong sign.  We
are grateful to Craig Roberts for bringing these errors to our
attention.  In the same paper, the lattice data for $\lambda_3$ and
for $\tau_5$ at the quark reflection point also had the wrong sign.}
\begin{gather}
\lambda_1^{\text{QED}}(p^2) = A(p^2) \, ; \label{eq:bc01} \\
\lambda_2^{\text{QED}}(p^2) = 
 -\half\frac{\mathrm{d}}{\mathrm{d}p^2}A(p^2) \, ; \label{eq:bc02} \\
\lambda_3^{\text{QED}}(p^2) =
 \frac{\mathrm{d}}{\mathrm{d}p^2}B(p^2) \, .\label{eq:bc03}
\end{gather}

\begin{figure}
\includegraphics*[width=\colw]{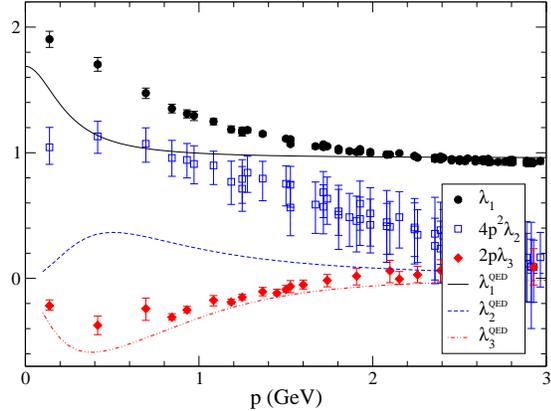}
\caption{The renormalised, dimensionless form factors $\lambda_1,
4p^2\lambda_2$ and $2p\lambda_3$ at the soft gluon point, as a
function of $p$, for $m=115$ MeV.  Also shown are the corresponding abelian
(Ball--Chiu) forms (\protect\ref{eq:bc01})--(\protect\ref{eq:bc03}),
derived from the quark propagator.}
\label{fig:q0}
\end{figure}

In Fig.~\ref{fig:q0} we show the dimensionless quantities $\lambda_1,
4p^2\lambda_2$ and $2p\lambda_3$ as a function of momentum $p$.  We also
show the abelian forms (\ref{eq:bc01})--(\ref{eq:bc03}) which have
been obtained from fitting lattice data for the quark propagator
\cite{Skullerud:2003qu,Bowman:2002kn}.  All these form factors have
been renormalised at 2 GeV, requiring
$\lambda_1(4\ \mathrm{GeV}^2,0,4\ \mathrm{GeV}^2)=1$.

We find that while $\lambda_3$ is quite close to its abelian form,
both $\lambda_1$ and $\lambda_2$ are significantly enhanced.  Since
the ghost self-energy would contribute the same prefactor (in this
kinematics, a constant) to all three form factors compared to the
abelian form, this points to a nontrivial structure in the
quark--ghost scattering kernel.  However, the singular nature of the soft gluon
point along with our small lattice volume make it
difficult to draw firmer conclusions.

\subsection{Quark reflection point $p=-k$}

\begin{figure}
\includegraphics*[width=\colw]{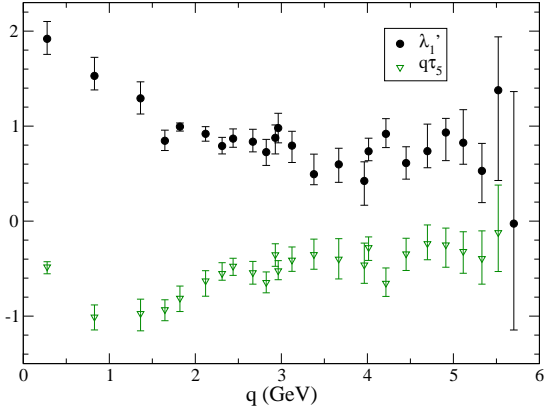}
\caption{The renormalised, dimensionless form factors $\lambda_1'$ and
$q\tau_5$ at the quark reflection point, as a function of the gluon
momentum $q$, for $m=115$ MeV.}
\label{fig:refl}
\end{figure}

At the quark reflection point $p=-k, q=-2p$ only $\lambda'_1$ and
$\tau_5$ survive, and the projected vertex is
\begin{equation}
\begin{split}
  \Gamma^P_\mu(-q/2,q) =& 
 \lambda'_1(q^2)\bigl(\gm-\qslash q_\mu/q^2\bigr) \\
& - i\tau_5(q^2)\smn q_\nu \, ,
\end{split}
\end{equation}
where in this section we write $\{\lambda'_1,\tau_5\}(q^2/4,q^2,q^2/4)
= \{\lambda'_1,\tau_5\}(q^2)$.  The dimensionless quantities
$\lambda'_1(q^2), q\tau_5(q^2)$ are shown in Fig.~\ref{fig:refl}.  These
form factors have been renormalised % at 2 GeV in the \MOMB\ scheme,
requiring $\lambda'_1(1\mathrm{GeV}^2,4\mathrm{GeV}^2,1\mathrm{GeV}^2)=1$.
$\lambda'_1$ shows the same qualitative behavour as $\lambda_1$ at the
soft gluon point, with a quite strong infrared enhancement.  We see
that $\tau_5$, which has rarely if ever been included in vertex models
used in DSE studies, is quite sizeable, indeed comparable in magnitude
to the dominant component $\lambda'_1$. It will be interesting to
study the effect of including this part of the vertex in future DSE studies.

\subsection{General kinematics}

The general lattice tensor structure, even for the Dirac-vector part of
the vertex alone, is very complicated and makes a full determination
of the vertex very difficult with this lattice action.  However, in
the special case where we choose both the quark and gluon momentum
vectors to be `perpendicular' to the vertex component, i.e.\ if we compute
$\Gamma_\mu(p,q)$ with $p_\mu=q_\mu=0$, this structure simplifies
considerably.  There is no loss of generality provided rotational
symmetry is restored in the continuum.   Here, we will only study the
leading, vector part of the vertex, but the other components may also
be determined in principle.  In continuum notation, we compute
\begin{align}
\begin{split}
\frac{1}{4}\Tr\gm&\Gamma^P_\mu(p,q) = 
 \Bigl(1-\frac{q_\mu^2}{q^2}\Bigr)\lambda'_1 \\
& + \frac{2}{q^2}\Bigl[(pq)k_\mu-(kq)p_\mu\Bigr](p_\mu+k_\mu)\lambda'_2 \\
& - [k^2-p^2-(k_\mu^2-p_\mu^2)]\tau_6
\end{split} \\
=\,& \lambda'_1 - (k^2-p^2)\tau_6 \equiv \lambda'' \, .
\end{align}
\begin{figure}
\includegraphics[width=\colw]{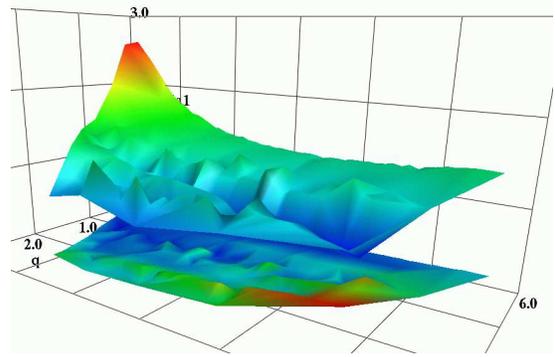}
\caption{The unrenormalised form factor $\lambda'_1$ in the
  quark-symmetric kinematics $p^2=k^2$ (upper surface), as a function of quark
  momentum $p$ (long axis) and gluon momentum $q$ (short axis) in
  units of GeV.  
  Statistical uncertainties are illustrated by the lower surface.}
\label{fig:qsym}
\end{figure}
Of particular interest is the quark-symmetric limit, where the two
quark momenta are equal in magnitude, $p^2=k^2$.  In this case,
$\tau_6$ is also eliminated, i.e.\ $\lambda''_1(p^2,q^2,p^2) =
\lambda'_1(p^2,q^2,p^2)$.  Note that both the soft gluon and the quark
reflection kinematics discussed previously are specific instances of
this more general case.  The details of the lattice implementation of
this, including the tree-level correction, will be described
elsewhere \cite{Skullerud:2004xx}.  In Fig.~\ref{fig:qsym} we show
$\lambda'_1$ as a function of the two remaining independent momentum
invariants. 
The data become quite noisy as $q$ is increased, and
also exhibit some `spikes' and `troughs' which at present we assume to
be numerical noise and lattice artefacts.

\begin{figure}
\includegraphics[width=\colw]{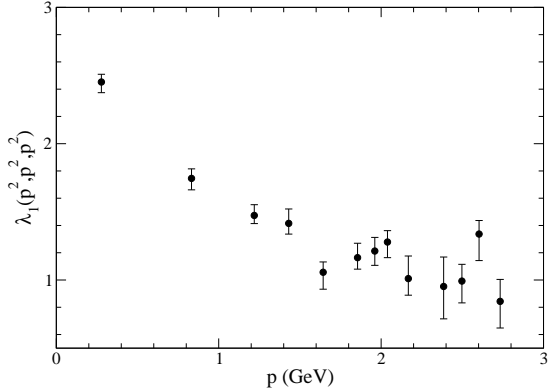}
\caption{The unrenormalised form factor $\lambda'_1$ at the totally
  symmetric kinematics $p^2=k^2=q^2$, as a function of the momentum
  $p$.}
\label{fig:symmetric}
\end{figure}

By interpolating the points in Fig.~\ref{fig:qsym}, we may reach the
totally symmetric point where $p^2=k^2=q^2$.  This kinematics has
a history of being used to define a momentum subtraction (MOM) scheme
\cite{Celmaster:1979km}.  We show our results in
Fig.~\ref{fig:symmetric}.  Again we find a strong infrared
enhancement.
%, although we have too few points available and the quality
%of the data is too poor to quantify this in any reliable way.
%

\begin{figure}[tbh]
\includegraphics*[width=\colw]{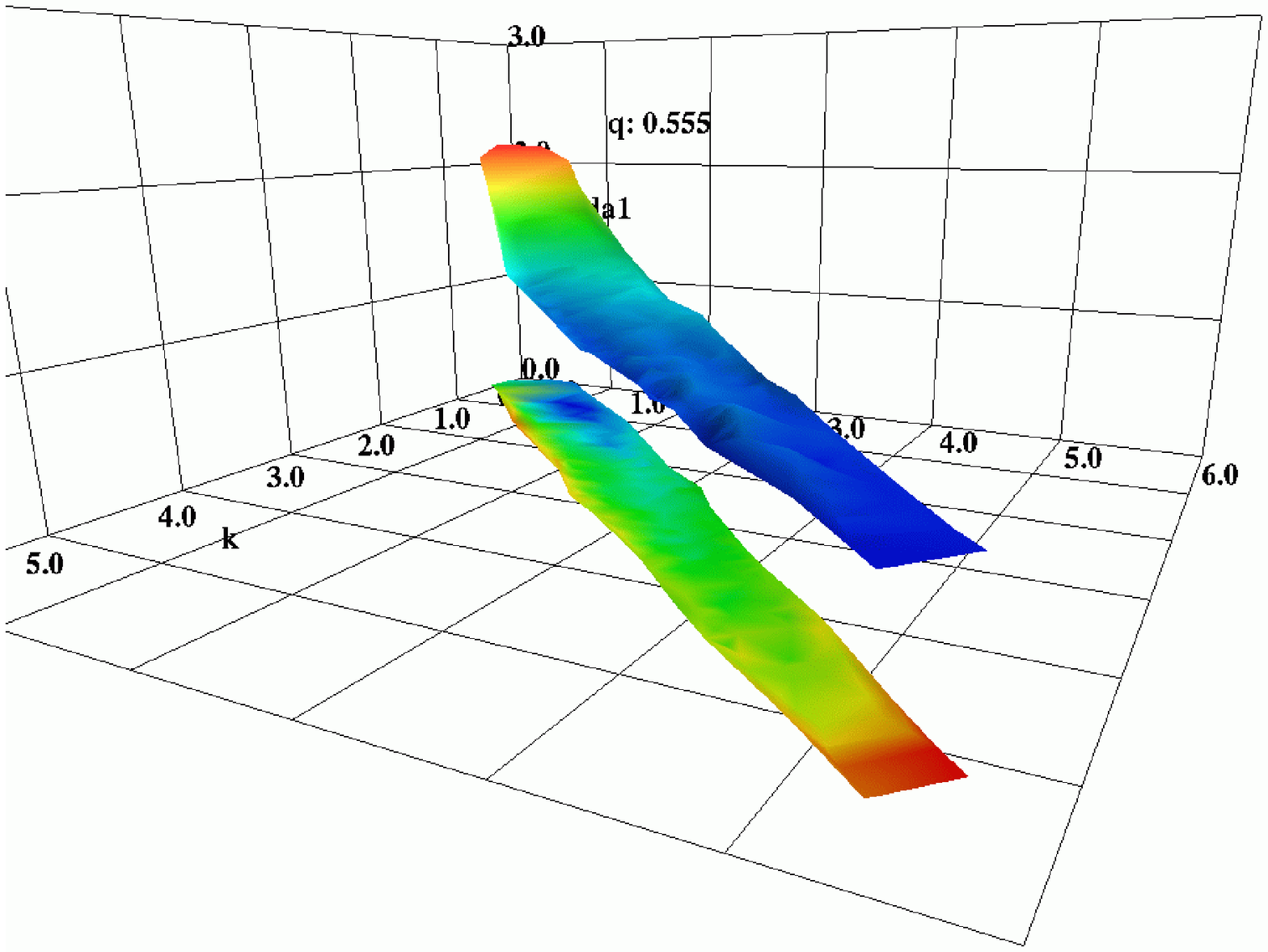}
\includegraphics*[width=\colw]{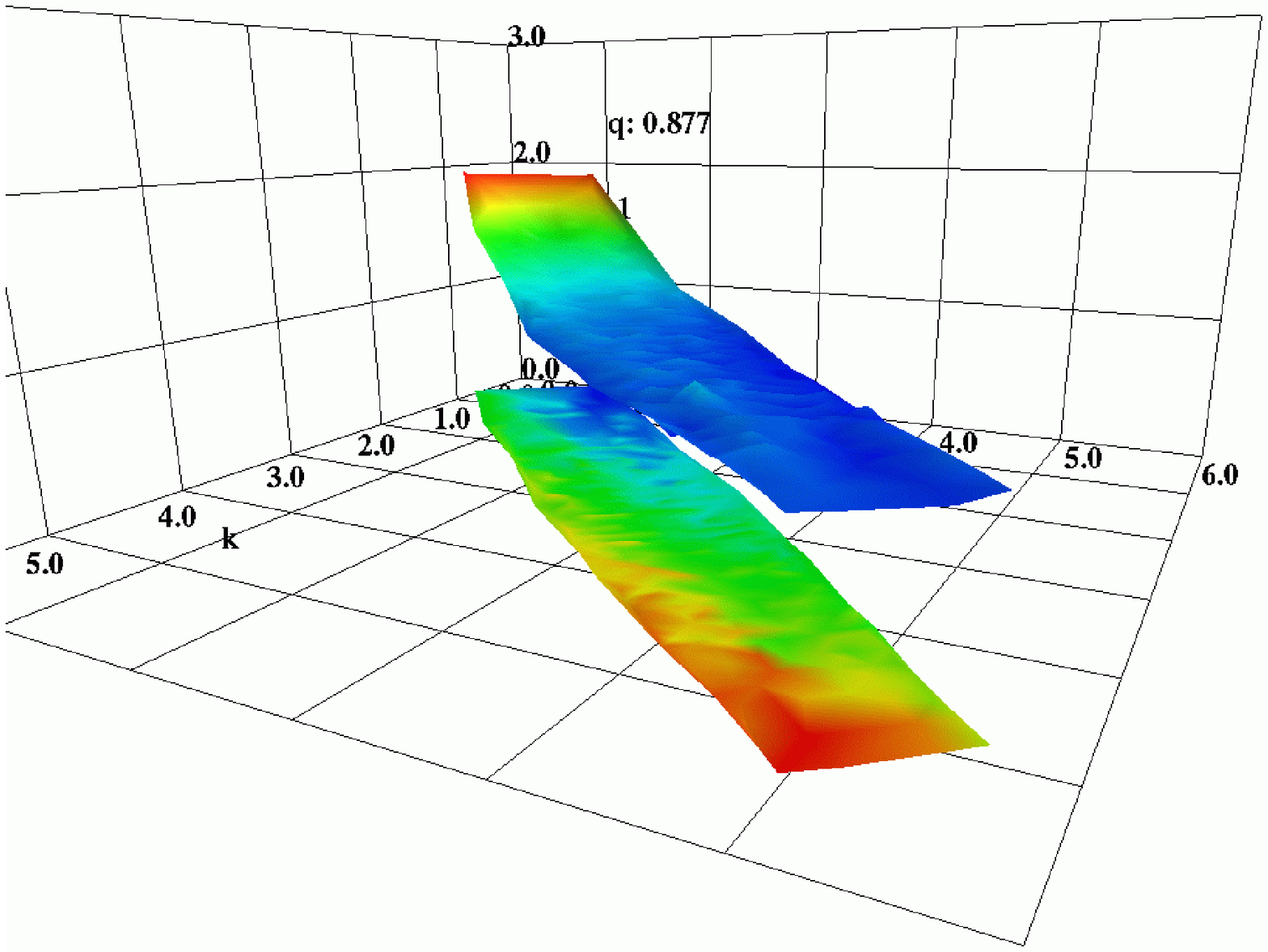}
\caption{The unrenormalised form factor $\lambda''_1$ for gluon
   momentum $q=0.555$ GeV (top) and $q=0.873$ GeV (bottom), as a
   function of quark momenta $p$ and $k$.  The lower surfaces denote
   the statistical uncertainties.}
\label{fig:asym1}
\end{figure}
\begin{figure}[tbh]
%\vspace{-1.0cm}
\includegraphics*[width=\colw]{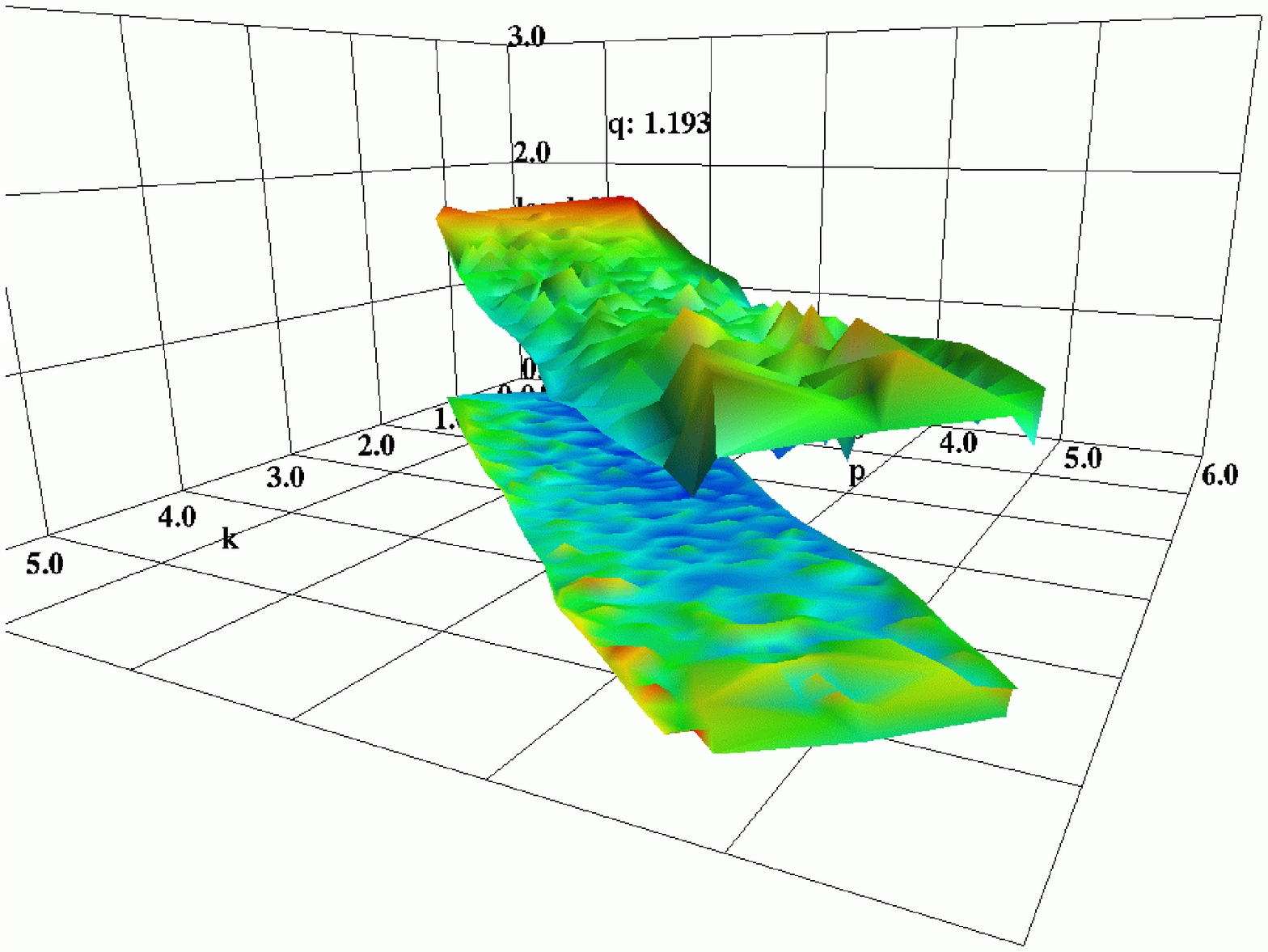}
\includegraphics*[width=\colw]{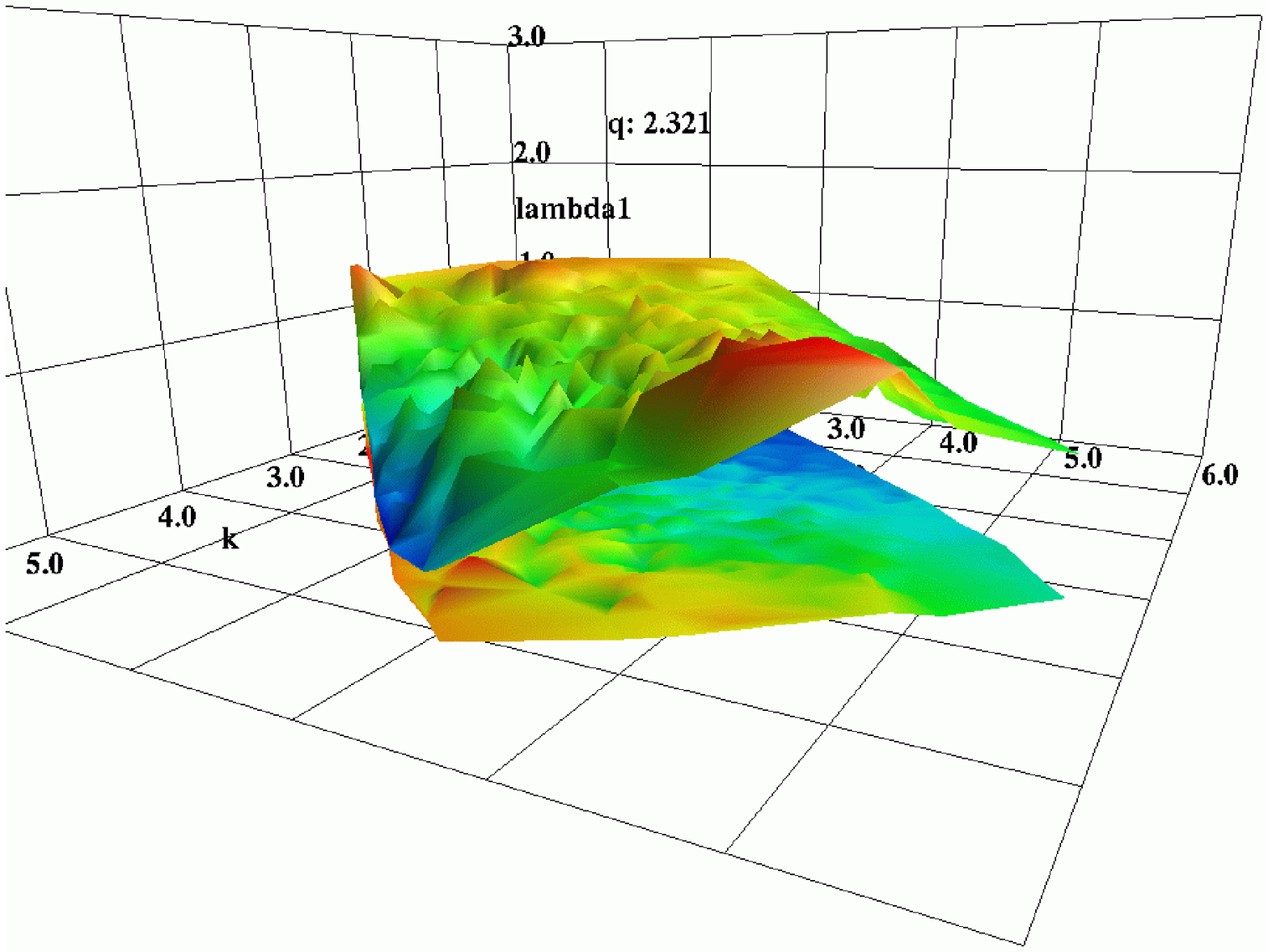}
%\vspace{-0.50cm}
\caption{As Fig.~\protect\ref{fig:asym1}, but for gluon 
   momentum $q=1.193$ GeV and 2.321 GeV.}
\label{fig:asym3}
%\vspace{-0.60cm}
\end{figure}
Finally, figs.~\ref{fig:asym1} and \ref{fig:asym3} show $\lambda_1''$
in general kinematics, for four different fixed values of $q$, as a
function of the two quark momenta $p$ and $k$.  We expect all form
factors to be symmetric in $p^2$ and $k^2$ ($\tau_6$ on its own is
antisymmetric, but is multiplied by $p^2-k^2$), and this is also what
the figures show, within errors.  The broadening of the data surface
as $q$ grows is simply a reflection of the increase in available phase
space.

The same qualitiative features as were found in the more specific
kinematics, are reproduced here.  At low $q$, we see a clear infrared
enhancement, which disappears as $q$ grows, reflecting the fact that
at high momentum scales, only the logarithmic behaviour (which is too
weak to be seen in these data) remains.  At the same time, the level
of the surface sinks, which reflects the infrared enhancement of
$\lambda_1''$ also as a function of gluon momentum.

\section{DISCUSSION AND OUTLOOK}

We have determined the complete tensor structure of the quark--gluon
vertex at two kinematical points, as well as the leading component in
arbitrary kinematics.  At the soft gluon point, we have observed
significant and non-uniform deviations from the abelian form which has
previously been the basis for DSE studies.  At the quark reflection
point, we find a significant contribution from the `chromomagnetic'
form factor $\tau_5$, which has previously been ignored.  In general
kinematics, we find an infrared enhancement in all momentum
directions; we hope to be able to quantify this enhancement more
clearly by fitting the data in the infrared region to functional forms
in the three momentum variables $(p^2,k^2,q^2)$.

It is interesting to compare these results with recent calculations
based on nonperturbative extensions of the one-loop vertex
\cite{Bhagwat:2004kj,Fischer:2004ym}. Both these studies agree very
well with our results for $\lambda_1$ and $\lambda_3$, while finding
substantially lower values for $\lambda_2$.  Since all these
calculations must be considered preliminary, too much emphasis should
not be placed on this.  It is worth noting that $\lambda_2$ is
inherently noisy, as it mixes with $\lambda_1$, which must be
subtracted in order to obtain the data shown in Fig.~\ref{fig:q0}.

These results have been obtained on a rather small lattice, and with a
discretisation that gives rise to quite large tree-level lattice
artefacts which must be corrected for.  We therefore expect systematic
errors to be large for large momenta.  To obtain more reliable results, and to
extend this study to the full vertex structure at all kinematics, it
would be desirable to employ an action which is known to have smaller
and more tractable tree-level artefacts.  The Asqtad action has been
employed successfully in computing the quark propagator
\cite{Bowman:2002bm}, and unlike the SW action, only $\lambda_1$ and
possibly $\lambda_2$ are non-zero at tree level, so tree-level
correction will not be needed for the remaining form factors.  This action
is also computationally cheap, making large lattice volumes
feasible.  Another possibility is to use overlap fermions, which have
the advantage of retaining an exact chiral symmetry, which
protects all the odd Dirac components of the vertex at tree level.

\section*{Acknowledgments}

This work has been supported by the Australian Research Council and
the Irish Research Council for Science, Engineering and Technology.
JIS is grateful for the hospitality of the Centre for the Subatomic
Structure of Matter, where part of this work was carried out.  We
thank Reinhard Alkofer, Christian Fischer and Craig Roberts for
stimulating discussions.

%\bibliography{lattice,gluon,qcd}

\begin{thebibliography}{10}

\bibitem{Bonnet:2001uh}
F.~D.~R. Bonnet, P.~O. Bowman, D.~B. Leinweber, A.~G. Williams and J.~M.
  Zanotti,
\newblock Phys. Rev. {\bf D64}, 034501 (2001) [hep-lat/0101013].
%%CITATION = HEP-LAT 0101013;%%

\bibitem{Bowman:2004jm}
P.~O. Bowman, U.~M. Heller, D.~B. Leinweber, M.~B. Parappilly and A.~G.
  Williams,
\newblock hep-lat/0402032.
%%CITATION = HEP-LAT 0402032;%%

\bibitem{Holl:2004qn}
A.~H{\"o}ll, A.~Krassnigg and C.~D. Roberts,
\newblock nucl-th/0408015;
%%CITATION = NUCL-TH 0408015;%%
%\bibitem{Bhagwat:2004hn}
M.~S. Bhagwat, A.~H{\"o}ll, A.~Krassnigg, C.~D. Roberts and P.~C. Tandy,
\newblock nucl-th/0403012.
%%CITATION = NUCL-TH 0403012;%%

\bibitem{Iida:2003xe}
H.~Iida, M.~Oka and H.~Suganuma,
\newblock hep-ph/0312328; H.~Iida, these proceedings.
%%CITATION = HEP-PH 0312328;%%

\bibitem{Fischer:2004ym}
C.~S. Fischer, F.~Llanes-Estrada and R.~Alkofer,
\newblock hep-ph/0407294.
%%CITATION = HEP-PH 0407294;%%

\bibitem{Tandy:2004rk}
P.~C.~Tandy,
%``Nonperturbative QCD Phenomenology and Light Quark Physics,''
\newblock nucl-th/0408037.
%%CITATION = NUCL-TH 0408037;%%

\bibitem{Bloch:2003sk}
J.~C.~R. Bloch, A.~Cucchieri, K.~Langfeld and T.~Mendes,
\newblock Nucl. Phys. {\bf B687}, 76 (2004) [hep-lat/0312036].
%%CITATION = HEP-LAT 0312036;%%

\bibitem{Fischer:2002hn}
C.~S. Fischer and R.~Alkofer,
\newblock Phys. Lett. {\bf B536}, 177 (2002) [hep-ph/0202202].
%%CITATION = HEP-PH 0202202;%%

\bibitem{Skullerud:2002ge}
J.~Skullerud and A.~K{\i}z{\i}lers{\"u},
\newblock JHEP {\bf 09}, 013 (2002) [hep-ph/0205318].
%%CITATION = HEP-PH 0205318;%%

\bibitem{Skullerud:2003qu}
J.~I. Skullerud, P.~O. Bowman, A.~K{\i}z{\i}lers{\"u}, D.~B. Leinweber and
  A.~G. Williams,
\newblock JHEP {\bf 04}, 047 (2003) [hep-ph/0303176].
%%CITATION = HEP-PH 0303176;%%

\bibitem{Bowler:1999ae}
UKQCD, K.~C. Bowler {\em et~al.},
\newblock Phys. Rev. {\bf D62}, 054506 (2000) [hep-lat/9910022].
%%CITATION = PHRVA,D62,054506;%%

\bibitem{Bowman:2002kn}
P.~O. Bowman, U.~M. Heller, D.~B. Leinweber and A.~G. Williams,
\newblock Nucl. Phys. Proc. Suppl. {\bf 119}, 323 (2003) [hep-lat/0209129].
%%CITATION = HEP-LAT 0209129;%%

\bibitem{Skullerud:2004xx}
J.-I. Skullerud {\em et~al.},
\newblock Quark-gluon vertex in general kinematics,
\newblock in preparation.

\bibitem{Celmaster:1979km}
W.~Celmaster and R.~J. Gonsalves,
\newblock Phys. Rev. {\bf D20}, 1420 (1979).
%%CITATION = PHRVA,D20,1420;%%

\bibitem{Bhagwat:2004kj}
M.~S. Bhagwat and P.~C. Tandy,
\newblock hep-ph/0407163.
%%CITATION = HEP-PH 0407163;%%

\bibitem{Bowman:2002bm}
P.~O. Bowman, U.~M. Heller and A.~G. Williams,
\newblock Phys. Rev. {\bf D66}, 014505 (2002) [hep-lat/0203001].
%%CITATION = HEP-LAT 0203001;%%

\end{thebibliography}

\end{document}